\documentclass[twocolumn,prl,amsmath,amssymb]{revtex4}
\usepackage{epsfig}
\usepackage{graphicx}
\usepackage{amssymb}
\usepackage{dcolumn}
\usepackage{color}
\usepackage{bm}
\usepackage{gensymb}
 
\begin{document}

\title{Sub 200fs pulse generation from a graphene mode-locked fiber laser}

\author{D. Popa, Z. Sun, F. Torrisi, T. Hasan, F. Wang, A. C. Ferrari}
\email{acf26@eng.cam.ac.uk}
\affiliation{Department of Engineering, University of Cambridge,Cambridge CB3 0FA, UK}

\begin{abstract} Ultrafast fiber lasers with short pulses and broad bandwidth are in great demand for a variety of applications, such as spectroscopy, biomedical diagnosis and optical communications. In particular sub-200fs pulses are required for ultrafast spectroscopy with high temporal resolution. Graphene is an ideal ultra-wide-band saturable absorber. We report the generation of 174fs pulses from a graphene-based fiber laser.

\end{abstract}
\maketitle

Ultrafast fiber lasers have many applications, ranging from basic research to materials processing and medicine\cite{Dausinger2004,Keller2003}. Compared to other types of lasers, they have a simple and compact design, efficient heat dissipation and high-quality pulse generation\cite{Okhotnikov2004}. The latter is typically achieved through a nonlinear optical device called saturable absorber (SA)\cite{Fermann1993,Brabec1992}. The SA technology is currently dominated by semiconductor SA mirrors\cite{Keller2003,Steinmeyer1999,Okhotnikov2004}. However these require complex and expensive fabrication and packaging and have limited bandwidth\cite{Keller2003,Steinmeyer1999,Okhotnikov2004}. Single wall carbon nanotubes (SWNTs) and graphene are ideal SAs, having fast recovery times, low saturation intensity, low cost and easy fabrication\cite{Geim,Wang2008,Sun2010g,Hasan2009,Set2004,DellaValle2006,Tausenev2008,Sun2008,Scardaci2008,Sun2009hp,sunnanores_10,Bonaccorso2010,Kivisto2009,Breusing2009,D.Sun2008,Kampfrath2005}. Broadband operation can be achieved using a distribution of tube diameters\cite{Wang2008}. However, this is an intrinsic property of graphene, due to the gapless linear dispersion of Dirac electrons\cite{Geim,Sun2010g,Sun2010tg,Bonaccorso2010,Casiraghi2007,Nair2008}.

Soliton-like mode-locking is the conventional method to generate ultrashort pulses in fiber lasers, down to several hundred fs\cite{Wang2008,Sun2010g,Set2004,Tausenev2008,sunnanores_10,Mollenauer1980}. In this regime, the pulse is shaped by the cavity dispersion and nonlinearity. Using this technique,$\sim$400fs pulses were reported for SWNT-based SAs\cite{Sun2008,Kivisto2009} and graphene-based SAs\cite{Sun2010g,Zhang2009}. The area theorem\cite{McCall1967} for solitons states that the product of pulse energy $(E)$ and duration ($\tau$) is fixed by the cavity dispersion and nonlinearity\cite{nelson1997}: $E\times\tau\propto \beta_{2}/n_{2}$, with $\beta_{2}$ the group velocity dispersion (GVD) coefficient and $n_{2}$ the nonlinear refractive index of the propagating medium. Thus, for a given system design, there is a trade-off between $E$ and $\tau$. One approach to mitigate these effects is to alternate segments of large normal (positive) and anomalous (negative) GVD fiber. In this way, the average pulse width in one cavity round trip can increase by an order of magnitude or more\cite{nelson1997}, significantly decreasing the intracavity average peak power, compared to soliton-like operation\cite{nelson1997}. This decreases nonlinear effects, reducing pulse duration\cite{Shohda2010}.

Here, we report a dispersion-managed fiber laser mode-locked by a graphene-based SA (GSA). The cavity comprises sections with positive and negative dispersion, provided by an erbium doped (EDF) and a single mode fiber (SMF). We get$\sim$174fs pulses with 15.6nm spectral width, much shorter than reported thus far for GSAs\cite{Sun2010g,Zhang2009}.
\begin{figure}
\centerline{\includegraphics[width=75mm]{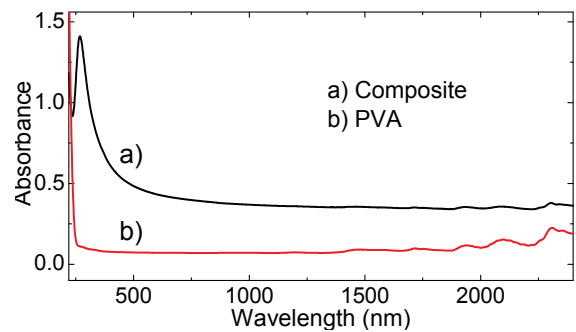}}
\caption{Absorption spectra of a) Composite and b) PVA.}
\label{abs}
\end{figure}
\begin{figure}
\centerline{\includegraphics[width=75mm]{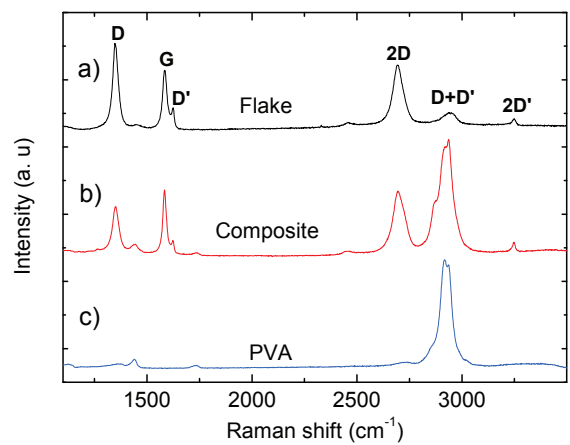}}
\caption{Raman spectra of a) Flake; b) Composite and c) PVA}
\label{raman}
\end{figure}

Graphite flakes are exfoliated by mild ultrasonication with sodium deoxycholate (SDC) surfactant\cite{Sun2010g,Hernandez2008}. A dispersion enriched with single (SLG) and few layer graphene (FLG) is then mixed with an aqueous solution of polyvinyl alcohol (PVA). After water evaporation, a graphene-PVA composite is obtained\cite{Hasan2009,Sun2010g}. Absorption measurements show a featureless spectrum from 500 to 2000nm, Fig.\ref{abs}. The UV peak is a signature of the van Hove singularity in the graphene density of states\cite{Kravets2010}. The PVA only shows significant absorption for shorter wavelengths\cite{El-Kader}. Fig.\ref{raman}a plots a typical Raman spectrum of an exfoliated flake on Si/SiO$_{2}$. Besides the G and 2D peaks, it has D, D', D+D' bands. The G peak corresponds to the E$_{2g}$ phonon at the Brillouin zone centre\cite{Ferrari2006}. The D peak is due to the breathing modes of sp$^{2}$ rings and requires a defect for its activation by double resonance (DR)\cite{Ferrari2006, Ferrari2000}. The 2D peak is the second order of the D peak. This is a single band in SLG, whereas it splits in multi-layer graphite\cite{Ferrari2006}, reflecting the evolution of the band structure. The 2D peak is always seen, even when no D peak is present, since no defects are required for the activation of two phonons with the same momentum, one backscattering from the other. DR can also happen intra-valley, {\it i.e.} connecting two points belonging to the same cone around \textbf{K} or \textbf{K'}. This gives rise to the D' peak. The 2D' is the second order of the D' peak. We do not assign the D and D' intensity in Fig.\ref{raman}a to the presence of a large amount of structural defects, otherwise they would be much broader, and G, D' would merge\cite{Ferrari2000}. We rather ascribe them to the edges of our sub-micrometer flakes\cite{Casiraghi2009}. Although 2D is broader than in pristine graphene, is still a single Lorentzian. Thus, even if the flakes are multi-layers, they behave, to a first approximation, as a SLG ensemble. Fig.\ref{raman}c is a reference PVA. The spectrum of the graphene-PVA composite is a superposition of Figs.\ref{raman}(a,c): embedding into PVA preserves the structure.
\begin{figure}
\centerline{\includegraphics[width=75mm]{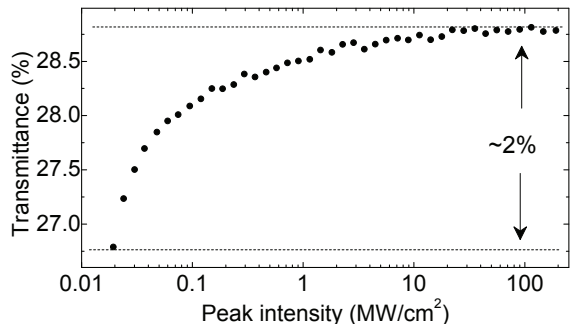}}
\caption{Transmittance as a function of peak intensity.}
\label{tra}
\end{figure}

Power-dependent absorbtion is measured with a SWNT-based mode-locked fiber laser, delivering 450fs pulses at 1558nm with 38MHz repetition rate\cite{Sun2008}. The signal is directed to the sample via an attenuator. A double-channel power meter measures both input and output power. The GSA is then placed between two fiber connectors. The output is monitored while the input power is varied, Fig.\ref{tra}. The maximum peak intensity $I_{peak}\sim$337MW/cm$^{2}$ is reached for an average pump power$\sim$5.15mW. The insertion loss is$\sim$5dB (i.e.$\sim$28\% transmittance), acceptable for the single-pass gain of a fiber laser\cite{Wang2008}. The transmittance increases by 2\% when the SA saturates at $I_{peak}\sim$100MW/cm$^{2}$. The modulation depth is 2\%, as in SWNT-SAs\cite{Wang2008,Scardaci2008,Sun2009hp}.
\begin{figure}
\centerline{\includegraphics[width=75mm]{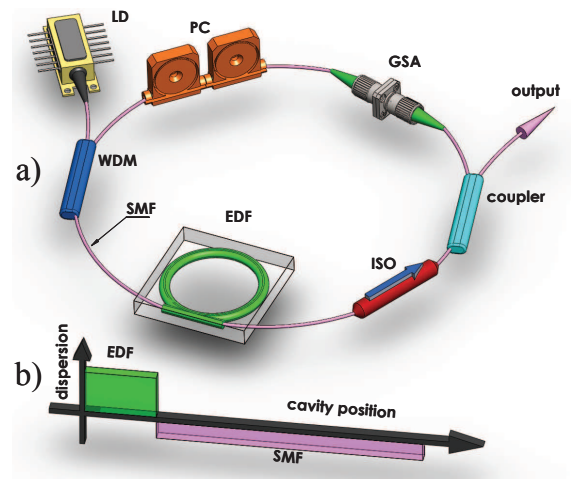}}
\caption{a) Laser setup. LD (laser diode), WDM (wavelength division multiplexer), SMF (single mode fiber), EDF (erbium-doped fiber), ISO (isolator), GSA (graphene saturable absorber), PC (polarization controller). b) Dispersion map. Positive and negative dispersion provided by EDF and SMF.}
\label{setup}
\end{figure}

Fig.\ref{setup}a is a schematic laser setup. We use 1.25m highly-doped EDF as gain medium, pumped by a 980nm laser diode (LD) through a fused wavelength division multiplexer (WDM). The laser output is directed through the 20\% port of a coupler. The total cavity length is$\sim$7.6m. We use a EDF with $\beta_{2}$=48ps$^{2}$/km, estimated by inserting it into the SWNT-mode-locked soliton-like fiber laser described above, and measuring the shift ($\Delta\lambda$) between sidebands and central wavelength of the soliton pulse spectrum\cite{dennis1994}. The rest of the cavity consists of a combination of SMF Flexcor 1060 and SMF-28 with anomalous GVD, as indicated in the dispersion map of Fig.4b. The measured total intracavity GVD is$\sim$-0.052ps$^{2}$, typical for dispersion-managed cavities\cite{nelson1997}. The laser operation is monitored with a second harmonic generation (SHG) autocorrelator and an oscilloscope. An optical spectrum analyzer with 0.07nm resolution probes the output.
\begin{figure}
\centering{\includegraphics[width=75mm]{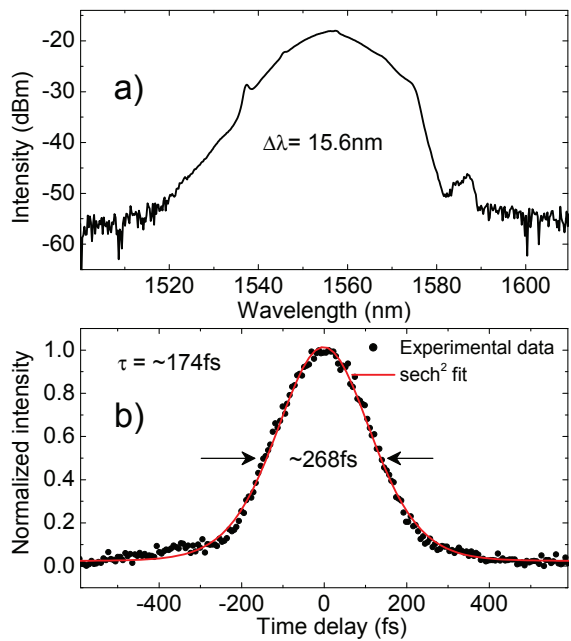}}
\caption{a) Optical spectrum with bandwidth $\Delta\lambda$=15.6nm. b) Autocorrelation trace with sech$^{2}$ fit.}
\label{spectra}
\end{figure}

Continuous wave (CW) operation starts at$\sim$18mW pump power, single-pulse mode-locking at$\sim$25mW. The repetition rate is 27.4MHz, as determined by the cavity length. The output power is$\sim$1.2mW for$\sim$28mW pump, with pulse energy$\sim$44pJ and $I_{peak}\sim$282MW/cm$^{2}$. Fig.\ref{spectra}a shows that the sidebands, expected for soliton operation, are small, as a result of temporal pulse stretching/compression\cite{tamura1994}. The full width at half maximum (FWHM) bandwidth is$\sim$15.6nm, larger than previous GSA-fiber lasers\cite{Sun2010g,Zhang2009,song2010,Bonaccorso2010}. Fig.\ref{spectra}b plots the SHG trace. For a sech$^{2}$ profile, we get a pulse duration$\sim$174fs, much shorter than previous GSA-mode-locked lasers\cite{Sun2010g,Zhang2009,song2010,Bonaccorso2010}. The time-bandwidth product is$\sim$0.335, slightly higher than 0.315, expected in the case of transform-limited sech$^{2}$ pulses, possibly due to uncompensated third-order dispersion distorting the intracavity pulse, thus limiting the minimum pulse width\cite{tamura1994}. To achieve even shorter pulses, SAs with higher modulation depth would be required\cite{Keller2003}. Higher power could be feasible by evanescent field interaction with the GSAs\cite{song2010,Bonaccorso2010}.

\begin{figure}
\centerline{\includegraphics[width=75mm]{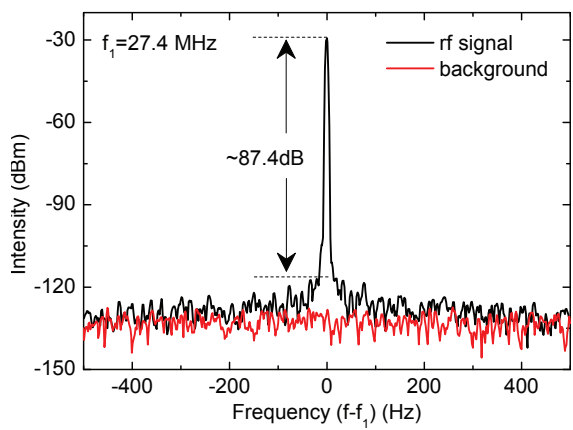}}
\caption{rf spectrum, measured around the fundamental repetition rate f$_{1}$=27.4MHz with 10Hz resolution.}
\label{rf}
\end{figure}

Stability is important for applications. We measured the radio frequency (rf) spectrum using a photodetector connected to a rf spectrum analyzer around the fundamental repetition rate (f$_{1}$=27.4MHz), Fig.\ref{rf}. The signal-to-noise ratio is$>80$dB, highlighting the low-amplitude fluctuations of our laser, thus stable mode-locking\cite{Linde1986}.

In conclusion, we demonstrated the use of a graphene-based SA as a mode locker for generation of pulses shorter than 200fs. The flexibility offered by the fiber laser design, along with the easy SA preparation, can lead to novel, low-cost ultrafast light sources.

We acknowledge funding from grants EPSRC-GR/S97613/01, EP/E500935/1, ERC-NANOPOTS, Royal Society Brian Mercer Award for Innovation and Wolfson Research Merit Award, CIKC in Advanced Manufacturing Technology, and King's College.

\end{document}